 \documentclass[final,twocolumn]{IEEEtran}
\usepackage{cite}
\usepackage{amsmath,amssymb,amsfonts}
\usepackage{algorithmic}
\usepackage{graphicx}
\usepackage{textcomp}
\def\BibTeX{{\rm B\kern-.05em{\sc i\kern-.025em b}\kern-.08em
		T\kern-.1667em\lower.7ex\hbox{E}\kern-.125emX}}

\usepackage{amsthm}
\usepackage{xargs}

 \usepackage[linesnumbered,ruled,vlined]{algorithm2e}

\SetKwInOut{Input}{Input}
 
\newcommand{\ket}[1]{|#1\rangle}
\newcommand{\bra}[1]{\langle#1|}

\newcommand{\op}[2]{| #1\rangle\!\langle #2 |}
\newcommand{\tr}{{\rm tr}}

\begin{document}

\title{Protocols for Packet Quantum Network Intercommunication}
\author{Nengkun Yu, Ching-Yi Lai, and Li Zhou
	\thanks{\footnotesize 
		NY is with the Centre for Quantum Software and Information, School of Software, Faculty of Engineering and Information Technology, University of Technology Sydney. (email: nengkunyu@gmail.com)
		
	CYL is with the Institute of Communications Engineering, National Yang Ming Chiao Tung University, Hsinchu 30010, Taiwan. (email:cylai@nycu.edu.tw)
	
	LZ was with the Department of Computer Science and Technology, Tsinghua University, Beijing, China.	
}}

\maketitle

\begin{abstract}
	A quantum network, which involves multiple parties pinging each other with quantum messages, could revolutionize communication, computing and basic sciences. The future internet will be a global system of various packet switching quantum and classical networks and we call it \emph{quantum internet}. To build a quantum internet, unified protocols that support the distribution of quantum messages
	within it are necessary.
	Intuitively one would extend classical internet protocols to handle quantum messages. However, classical network mechanisms, especially those related to error control and reliable connection, implicitly assume that information can be duplicated, which is not true in the quantum world due to the no-cloning theorem and monogamy of entanglement.
	In this paper, we investigate and propose protocols for packet quantum network intercommunication.
	To handle the packet loss problem in transport, we propose a quantum retransmission protocol based on the recursive use of a quantum secret sharing scheme. Other internet protocols are also discussed. 
	In particular, the creation of logical process-to-process connections   is accomplished by a quantum version of the three-way handshake protocol.
\end{abstract}

\section{Introduction}

Quantum computing is the next generation of computing that promises extraordinary computing power. Small-scale quantum processors have been constructed and continuously improved.
We are expecting a future with functional quantum processors around us.
With the additional ability to send quantum messages from one quantum processor to another, several quantum processors can form a quantum computing cluster and they can together build an entangled quantum system~\cite{VMD16}.
This is often referred  to as \emph{networked quantum computing}, or
\emph{distributed quantum computing}~(see, e.g., \cite{Raz99,RK11}),
{which may provide \emph{exceptional} savings in
	communication complexity compared with classical distributed computation.}
Together with   connected quantum computers,  the quantum internet   could also provide functionalities, such as secure communication with quantum key distribution~\cite{BB84,Bennett1992}, delegated quantum computation~\cite{BFK09,DSS16,Mah18}.

The concept of quantum internet was proposed decades ago~\cite{Elliott_2002,LSW+04} and a road map for the future quantum internet can be found, e.g., in \cite{WER18}.
Thanks to the rapid experimental progress in recent
years, first rudimentary quantum networks are not out of reach anymore.
Physicists can control and manipulate quantum signals much better than before~\cite{Kim08,RMP15}.
Very recently, the design
of quantum internet has received much attention from an engineering perspective. Many challenges are formulated in \cite{Caleffi:2018:QIC:3233188.3233224,CCT18}. 
{Generally, communication between two nodes in a quantum network is achieved by quantum teleportation~\cite{BBC+93}, where shared quantum entanglement is necessary.
	 Consequently, efficient and reliable protocols for routing entanglement  is vital to a quantum network \cite{P+19}.
	If two nodes are far from each other, additional nodes (called \emph{quantum repeaters}~\cite{BDCZ98}) serve as relays for quantum teleportation.   }
Basic design principles of such a quantum repeater network have been discussed in~\cite{MT13}.
Protocols for operations across layers of a quantum network  has also been discussed \cite{PD19}.
 Recently, Dahlberg et al. provided
a reliable physical and link layer protocol for quantum networks
on the NV hardware platform  \cite{DSC+19}.  A concurrent entanglement routing protocol for the 
the network layer is  proposed in \cite{SQ20}.
Several quantum network simulators are available~ (e.g.,~\cite{ABC+21,CKD+21}), facilitating the study of quantum networks.
 Taking the development of the classical internet for reference, the next step toward a future quantum internet is to establish reliable qubit transmission between quantum nodes. As quantum decoherence is inevitable in a quantum network~\cite{RVM12,KDW20}, this motivates us to study the quantum version of reliable \emph{transmission control protocol}.

In the classical internet, Transmission Control Protocol and Internet Protocol~(TCP/IP)~\cite{CK74} are the foundational protocols that serve as a  methodology of unified, reliable, ordered,
and error-checked delivery of   classical information stream between
applications   in the internet.
One would expect similar quantum analogues   that allow quantum computers on different platforms 
to interconnect.
However, the frames of classical TCP/IP cannot be directly applied in the quantum network because of the dramatic difference between classical bits and quantum bits (qubits).
In particular, retransmission is one of the basic mechanisms used in a packet switched   network for reliable communication. {However, retransmission of quantum messages is generally impossible due to the no-cloning theorem~\cite{WZ82}.
	Other instances such as \emph{checksum}, \emph{handshake} are also difficult.

	In this paper, we describe a future quantum internet model  and  study protocols for packet quantum network intercommunication in theory.
	The quantum internet model under consideration is the repeater-based model, where quantum communication is done by a serial of teleportations with quantum repeaters used as intermediate nodes to reach long distances.
	Packet switching is a technique for reliably and efficiently transmitting data in network communication and we will employ this feature in the quantum internet. Therefore, a quantum message will be divided into packets, which are then sent to the destination through a network of routers and repeaters. 
	Since quantum channels are inevitably noisy, it is important that the packets are protected by quantum error-correcting codes~\cite{PhysRevA.52.R2493}.
	In particular, the creation of  logical process-to-process connections  is accomplished by a quantum version of the \textit{three-way handshake protocol} 
	(see Section~\ref{sec:qTCP}).

	The most important issue of packet switching is~\textit{packet loss}.
	 EPR pairs shared between any nodes are constantly generated and distributed.
	However, entanglement can never be 100\% reliable no matter what accuracy  entanglement distillation and distribution can achieve. 
	There is no perfect quantum memory as well.
	If an imperfect EPR pair is used, a packet loss occurs.
	Other issues may also lead to packet loss, such as quantum decoherence, imperfect operations, nonavailability of fresh EPRs.
	Classically one would have the packet retransmitted but this is forbidden by the no-cloning theorem.
	Naively, one would like to use quantum error-correcting codes to protect the quantum message
	since   a packet loss could be considered as an erasure error. 
	However, quantum codes also cannot protect the information if there are too many errors beyond the error-correction capability.
	Herein we propose a quantum retransmission protocol (Protocol~\ref{prot:retransmission})
	by recursively using a (2,3) quantum secret sharing scheme (which is a variant of quantum codes).
	A quantum message will be ``divided" into three shares and any two shares suffice to recover the message.
 If a share gets lost, then we recover the share if there are two remaining shares or we divide the remaining share into another three shares and repeat the process. In this way, one can theoretically reconstruct the quantum message. One can show that by calculation only a constant number of shares are required to send a quantum message if the packet loss rate is low enough.

	The paper is organized as follows. Basics of quantum mechanics are given in the next section.
	The model of quantum internet is proposed in Section~\ref{sec:model}, including the protocols for entanglement distribution and router's action. In Section~\ref{sec:qTCP}, we propose the quantum three-way handshake protocol and the quantum retransmission protocol.
	Then we conclude.

\section{Preliminaries}
Here we present the bare-bones of quantum
mechanics for our purpose.
For more details, interested readers are referred to \cite{NC00}.
{We start with the postulates  of quantum mechanics.}
Then we introduce the effects of quantum teleportation, entanglement swapping,  the no-cloning theorem, monogamy of entanglement, and quantum error correction.

\subsection{Quantum Mechanics}

The state space of a closed quantum system is a complex Hilbert space
and a \emph{pure} quantum state is an arbitrary unit vector in the Hilbert space.
In particular,
a qubit system has a two-dimensional vector space with an orthonormal
basis $\{\ket{0},\ket{1}\}$.
The state of  a composite system of two subsystems is the tensor product of the state spaces of the two subsystems.  Hence 
an $n$-qubit system has a $2^n$-dimensional state space with
an orthonormal basis $\{ \ket{i_1i_2 \cdots i_n}\}  \triangleq \ket{i_1}\otimes \cdots \otimes \ket{i_n}:i_j\in\{0,1\}  \}$.
Note that $\otimes $ is the tensor product and will sometimes be omitted with no ambiguity.
More generally, a quantum state can be represented by a density operator $\rho$, which is
positive semi-definite and has trace equal to one.
For a pure quantum state $\ket{\psi}$,  its density operator is $\op{\psi}{\psi}$.
Quantum mechanics allows a more complicated state: a \emph{mixed} quantum state,
which is a  combination of some pure states $\rho=\sum_i p_i \ket{\psi_i}\bra{\psi_i}$
such that $p_i\geq 0$ and $\sum_{i}p_i=1$.

A two-qubit pure state is entangled if it cannot be described by two independent single-qubit states.
The state $\ket{\Phi^+}_{AB}=\frac{\ket{0}_A\ket{0}_{B}+\ket{1}_A\ket{1}_{B}}{\sqrt{2}}$ is called a \emph{maximally-entangled state}, or simply \emph{EPR pair}~\cite{PhysRev.47.777}, where the subscript $AB$ means that it is shared between $A$ and $B$.

The evolution of a closed quantum system can be described by a \emph{unitary} transformation.
An operator $U$ is  unitary if $U^{\dag}U=UU^{\dag}=I$, where $U^{\dag}$ is the complex conjugate transpose of~$U$, and $I$ is the identity operator.
A basis for the linear operators on a qubit is the Pauli matrices $I_2=\begin{bmatrix}1 &0\\0&1\end{bmatrix}, X=\begin{bmatrix}0 &1\\1&0\end{bmatrix},  Z=\begin{bmatrix}1 &0\\0&-1\end{bmatrix}, Y=iXZ$.

A general quantum operation, usually denoted by $\mathcal{E}$, is a completely positive and trace-preserving (CPTP) map on a density operator.
Consequently, a quantum communication channel is modeled as a quantum CPTP map.
In other words, if the sender sends   $\rho$ through a quantum channel $\mathcal{E}$, the receiver would receive   $\mathcal{E}(\rho)$.

A quantum measurement is a special quantum operation described by a collection of measurement operators $\{M_m\}$,
which satisfy $\sum_{m}M^{\dag}_mM_m=I,$
and the index $m$ stands for the measurement outcome.
If a  state $|\psi\rangle$ is measured,
the probability of outcome $m$ is
$p(m)=\langle \psi|M_m^{\dag}M_m|\psi\rangle$ and the post-measurement state is
$|\psi_m\rangle=\frac{M_m|\psi\rangle}{\sqrt{p(m)}}.$

\subsection{Quantum teleportation and entanglement swapping}
Quantum teleportation \cite{BBC+93} is arguably the most famous
quantum communication protocol. With the help
of pre-shared entanglement between the sender
and the receiver, quantum
information
can be transmitted from one location to another using
only classical communication.
The teleportation protocol is as follows.
Suppose Alice and Bob share an EPR pair $\ket{\Phi^+}_{AB}=\frac{\ket{0}_A\ket{0}_{B}+\ket{1}_A\ket{1}_{B}}{\sqrt{2}}$
and Alice wants to send to Bob an
unknown qubit
$
\ket{\psi}_C=\alpha\ket{0}_C+\beta\ket{1}_C.
$
She performs a \emph{Bell measurement} $\{M_{ij}\}$ on her two qubits $AC$,
where $$M_{ij}=  \ket{\Phi_{ij}}_{AC}\bra{\Phi_{ij}}_{AC},\quad  \ket{\Phi_{ij}}\triangleq (I_2\otimes X^i Z^j) \ket{\Phi^{+}}$$ for $i,j\in\{0,1\}$.
Alice then sends Bob her measurement outcome $ij\in \{00, 01, 10, 11\}$.
Interestingly, this two-bit message contains all the information
that Bob needs to recover $\ket{\psi}$ on his side.
To see this,   observe that
\begin{align*}
\ket{\Psi}_C\otimes\ket{\Phi^{+}}_{AB}=\frac{1}{2}\sum_{i,j\in\{0,1\}} \ket{\Phi_{ij}}_{AC}\otimes \left( X^iZ^j \ket{\psi}\right).
\end{align*}
According to Alice's measurement outcome $ij$, Bob's qubit will be in the state $X^{i}Z^{j} \ket{\psi}$
and thus $\ket{\psi}$ can be recovered after a \emph{Pauli correction} $X^{i}Z^{j}$ on Bob's qubit.

To sum up, with the help of an EPR pair, the transmission of two classical bits is enough to transmit one qubit.
Thus, one can send $n$ qubits by transmitting $2n$ classical bits using $n$ pre-shared EPR pairs.

Next we discuss a variation of quantum teleportation--\emph{entanglement swapping}~\cite{ZZHE93}.
Suppose that Alice and Bob share an EPR pair $\ket{\Phi^+}_{AB_1}$, and Bob and Charlie share another EPR pair $\ket{\Phi^+}_{B_2C}$. It is possible to construct an EPR pair shared between Alice and Charlie.
The procedure is as follows: Bob performs a Bell measurement on Bob's two qubits and sends the two-bit outcome to Charlie, who then performs a Pauli correction according to Bob's measurement outcome.
This can also be regarded as Bob teleporting his particle $B_1$ to Charlie by consuming the ERP pair $\ket{\Phi^+}_{B_2C}$. In other words, quantum correlations can be teleported.

\subsection{No-cloning theorem and monogamy of entanglement}

A fundamental property of quantum mechanics is that learning  an unknown quantum state
from a given specimen would disturb its state~\cite{BBJMPSW94}. In particular, the quantum no-cloning theorem \cite{DIEKS1982271,WZ82} states that an arbitrary unknown quantum state
cannot be cloned. Generally speaking,  there is no quantum operation that can transform an unknown quantum state $\ket{\psi}\otimes\ket{0}$ into $\ket{\psi}\otimes\ket{\psi}$.

Monogamy of entanglement, one of the most fundamental properties of entanglement, roughly says that the amount of shared entanglement among multiparties is bounded.
For example, suppose qubits A and B maximally-entangled, and then qubit C must be uncorrelated with A or B.

\subsection{Quantum error correction}
Qubits are  error-prone.
A quantum error-correcting code (QECC), first proposed by Shor~\cite{PhysRevA.52.R2493},  can protect quantum information against decoherence~\cite{NC00}.
Suppose we have a  noisy quantum channel denoted as a quantum operation $\mathcal{N}$.
If there exist encoding and decoding   operations $\mathcal{E}$ and  $\mathcal{D}$ such that
$$
(\mathcal{D}\circ\mathcal{N}\circ\mathcal{E})(\rho)\propto\rho,
$$
for any input state $\rho$, we say that $\mathcal{E}$ and $\mathcal{D}$ correct the errors of $\mathcal{N}$.
That is, any quantum information encoded by $\mathcal{E}$ can be perfectly recovered from the noise process $\mathcal{N}$ by applying the recovering map $\mathcal{D}$.

Quantum secret sharing schemes have close connection to  QECCs~\cite{CGL99}.
A $((k,n))$ threshold scheme with $n<2k$ encodes an arbitrary secret quantum state and divides it into $n$ shares such that
a) from any $k$ or more shares the secret can be recovered and b) from any $k-1$ or fewer shares no information about the secret quantum state can be deduced.
Note that the requirement of $n<2k$ comes from the no-cloning theorem.

\section{Quantum Internet Models} \label{sec:model}

\begin{figure}[t!]
	\includegraphics[width=7.5cm]{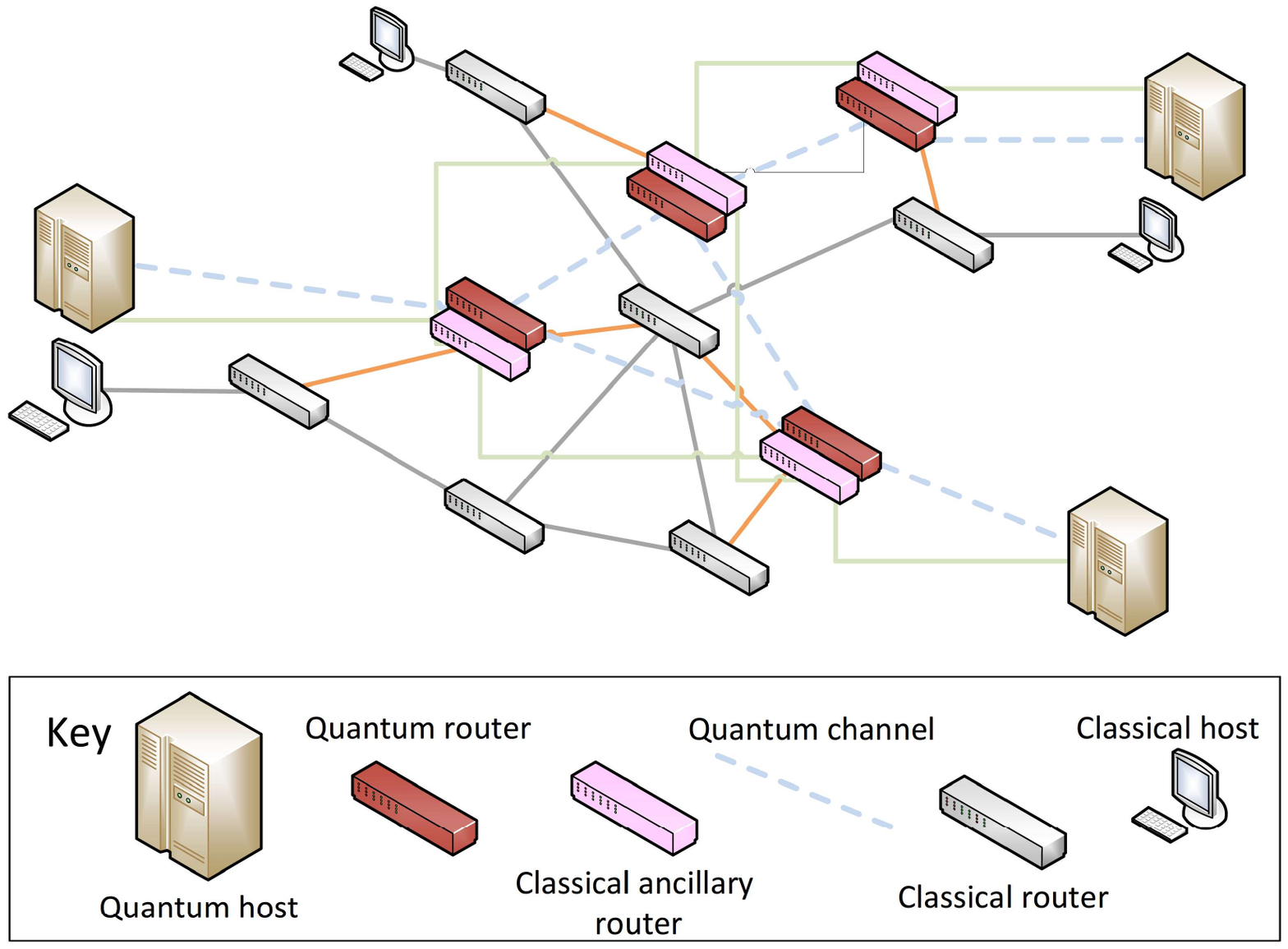}
\caption{Quantum Internet.} \label{fig:quantum_internet}
\end{figure}

Figure~\ref{fig:quantum_internet} illustrates the concept of a quantum network, where
several quantum computers are distributed and connected along with a classical network.
{PCs},  {workstations},  {Web servers}, quantum computers, etc, are called \emph{hosts}, or \emph{endnodes}.
Endnodes are connected together by a network of \textit{communication links} and
{routers}.

In this paper, we only discuss the transmission of quantum information through the internet.
We further assume that all the nodes within the quantum internet can exchange classical information, for example, over the classical internet.
{Since quantum teleportation  has distance constraints,  quantum repeaters are used as intermediate nodes.}
It is clear that EPRs shared between two neighboring nodes has to be constantly generated through the whole internet.
In the following we discuss the repeater-based quantum internet and the packet loss error model.

\subsection{Repeater-based Quantum Internet}

A quantum channel between two neighboring quantum devices is required for the transmission of qubits.
However,  quantum channels are inherently lossy and they are not reliable for long-distance communication.
Consequently, the techniques of \emph{quantum repeaters}
are used as intermediate nodes to reach long distances~\cite{PhysRevA.79.032325,PhysRevLett.98.190503,7010905,RevModPhys.83.33}.
\begin{figure}[h]
	\[	\includegraphics[width=7.cm]{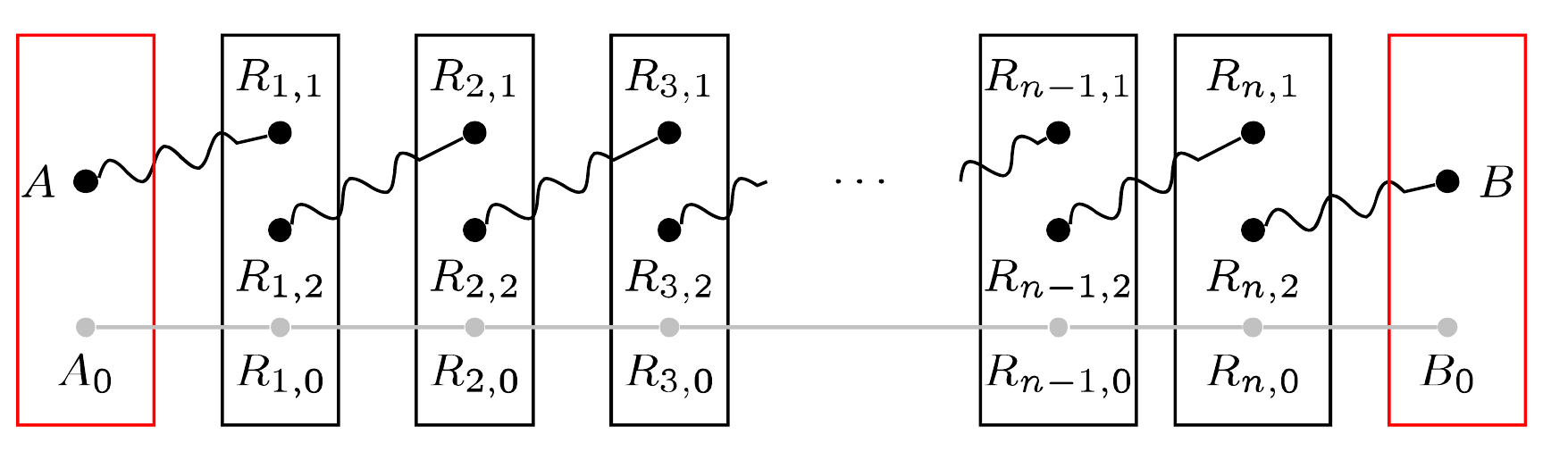} \]
	\caption{A repeater-based quantum channel.
		The gray lines are classical channels. Two entangled nodes are connected by a wavy line.  } \label{fig:quantum_repeater}
\end{figure}
The idea is to do a sequence of entanglement swapping or teleportation between two consecutive nodes
so that quantum communication between two endnodes  can be implemented as shown in Fig.~\ref{fig:quantum_repeater}.
Suppose two  {neighboring quantum devices} $A$ and $B$ are connected by repeaters  $R_1,\dots,R_n$, each of which has two (or more) qubits.  EPR pairs are constantly created between repeaters $R_{i,2}$ and $R_{i+1,1}$ for $i=1,\dots,n-1$, and between $A$ and $R_{1,1}$, $B$ and $R_{n,2}$.
Then each repeater performs a Bell measurement on its two qubits
and passes the measurement outcomes to $B$ for Pauli correction, using classical channels (the gray lines in Fig.~\ref{fig:quantum_repeater}).
Thus $A$ and $B$ can share an EPR pair.\footnote{This may be replaced by a process with deferred measurements, followed by an overall Pauli correction in $B$~\cite{NC00}.}

A delicate step of a repeater-based quantum channel is that the Pauli correction for teleportation
can be deferred.  {For example, suppose that a qubit is to be sent from node $A$ to $B$ via a router $R$.}
Originally, $A$ sends $R$ the binary measurement outcome $m_1$ and then $R$ does a Pauli correction according to $m_1$. Instead, $R$ can perform a Bell measurement on his qubits and send the binary outcome $m_2$, together with $m_1$, to $B$ for Pauli correction. In fact, having $m_1+m_2 \mod 2$ is enough for $B$ to recover the transmitted qubit.

\subsection{Quantum Packets and Packet Loss}

Classically, whenever one party sends certain data to another party, it will retain a copy of the data until that the recipient has acknowledged receipt of the data. 
A lost packet can be retransmitted from the sender using the retained copy.

In a quantum network, quantum communication may be
disrupted due to unrecoverable mutation of the data
or missing data. The reasons include quantum decoherence, imperfect operations, the network packet loss, and others.
End-to-end restoration procedures
are desirable to allow complete recovery from these
conditions.

However, quantum laws prevent us from using certain classical techniques. 
For example,
assume that a bipartite state $\ket{\psi}_{A_1A_2B}$ is shared between A and B, where $A_1A_2$ are held by A, and A wants to send $A_2$ to B. The following quantum features prohibit ideas from classical networks.
\begin{enumerate}
	\item No-cloning: In general, A   can not make a copy of $A_2$, says $A_2'$ while sending $A_2$ through the network.
	\item Monogamy: Even if A knows exactly the whole state $\ket{\psi}_{A_1A_2B}$ is a pure entangled state, A can not make a copy $A_2'$ such that the state of $A_1A_2'B$ is exactly $\ket{\psi}_{A_1A_2B}$ while sending $A_2$ through the network. Also, if $A_2$ is lost, B can not make a $B_2$ such that the state of $A_1B_2B$ is exactly $\ket{\psi}_{A_1A_2B}$, even if B knows the exact characterization of state $\ket{\psi}_{A_1A_2B}$.
\end{enumerate}

We will imitate the packet switching technique
and propose the quantum analogue for the quantum internet so that all quantum data is eventually transferred between each source-destination pair in the next section.

\section{Transport Layer: Quantum Transmission Control Protocol} \label{sec:qTCP}

Quantum information is fragile through the transmission over the internet.
Using quantum error-correcting codes to protect the packets is not enough for recovering the underlying quantum message.
To guarantee datagram delivery, a quantum version of information retransmission is needed regardless   the no-cloning theorem.
Herein we show how information retransmission can be achieved using the techniques of quantum secret sharing~\cite{CGL99}.
This guarantees that the quantum data stream transmitted through quantum Transmission Control Protocol (qTCP) will have exactly the same quantum information and correlation
as the original stream with high probability.

The qTCP   operations may be divided into three phases:
connection establishment, data transfer, and correction termination.
First, logical process-to-process connections 
are established by a quantum version of the
\emph{three-way handshake} protocol.
Next data streams are transferred; a \emph{retransmission protocol}  is used to prevent quantum packets from loss. Finally, a termination protocol closes established virtual circuits and releases all allocated resources.
To terminate a connection, we can use the four-way handshake as in classical TCP, where each side of the connection is terminated  and each buffer is released independently.  

Note that  each packet will be protected by a quantum error-correcting code and a quantum message will be split into several shares (packets) by a quantum secret sharing scheme. Consequently,  the quantum message is protected by a concatenation of a quantum code and a secret sharing scheme.

In the following subsections, we discuss the quantum three-way handshake and retransmission protocols, respectively.
We first introduce some tools from quantum error detection.

\subsection{Quantum error correction}
The widely-used error detection method, including  parity bits, Cyclic Redundancy Check, and  Checksum, can be characterized by a \emph{check function} $f:\{0,1\}^n\mapsto \{0,1\}^k$ as in the following encoding procedure:  a given $n$-bit string $s$  is encoded as $(s,f(s))$ of $(n+k)$ bits.
{This idea has been generalized to QECCs~\cite{NC00}.

Suppose  a host  wants to send  an $n$-qubit register $A$ through the internet, using a check function $f$.
Let $\ket{\psi}_{A,R}$ be a purified state of $A$ for   a reference system $R$.
The encoding is done by firstly appending  $k$ ancilla qubits in $\ket{0^k}_S$ to $A$, and  the state of $AS$ is of the form 
$$\sum_{0\leq j\leq 2^n-1}\alpha_{j}\ket{j}_{A}\ket{\phi_{j}}_R\ket{0^k}_S.$$
Then the host applies a unitary encoder $U$ on $AS$ to obtain
$$\sum_{0\leq j\leq 2^n-1}\alpha_{j}\ket{\bar{j}}_{AS}\ket{\phi_{j}}_R,$$
where $\ket{\bar{j}}_{AS}$ is the encoded $\ket{j}_A$.
Upon receiving $A S$, the receiver simply applies a decoding operation to $AS$ so that the state becomes
$$\sum_{0\leq j\leq 2^n-1}\sum_{\textbf{a},\textbf{b}}\alpha_{j}E_{\textbf{a},\textbf{b}}\ket{j}_{A}\ket{\phi_{j}}_R\ket{\textbf{a},\textbf{b}}_S,$$
where $\textbf{a},\textbf{b}$ denote the error syndrome for Pauli $X$ and $Z$, and  $E_{\textbf{a},\textbf{b}}$ is a corresponding error operator. If there is no logical error, measuring system $S$ in the computational basiswould tell us an appropriate recovery operation and we can recover the original state.
It could be the case that some error occurs but the measurement outcome is $0^k$,
which will lead to a decoding error. This situation occurs with a small chance if the QECC is chosen appropriately.}

In the following, a packet will be protected by such an error-correcting code. This allows us to justify whether a packet is valid or not.

\subsection{Connection establishment }

Figure~\ref{fig:3} illustrates the idea of quantum three-way handshake protocol. 
\begin{figure}\[
	\includegraphics[width=7.5cm]{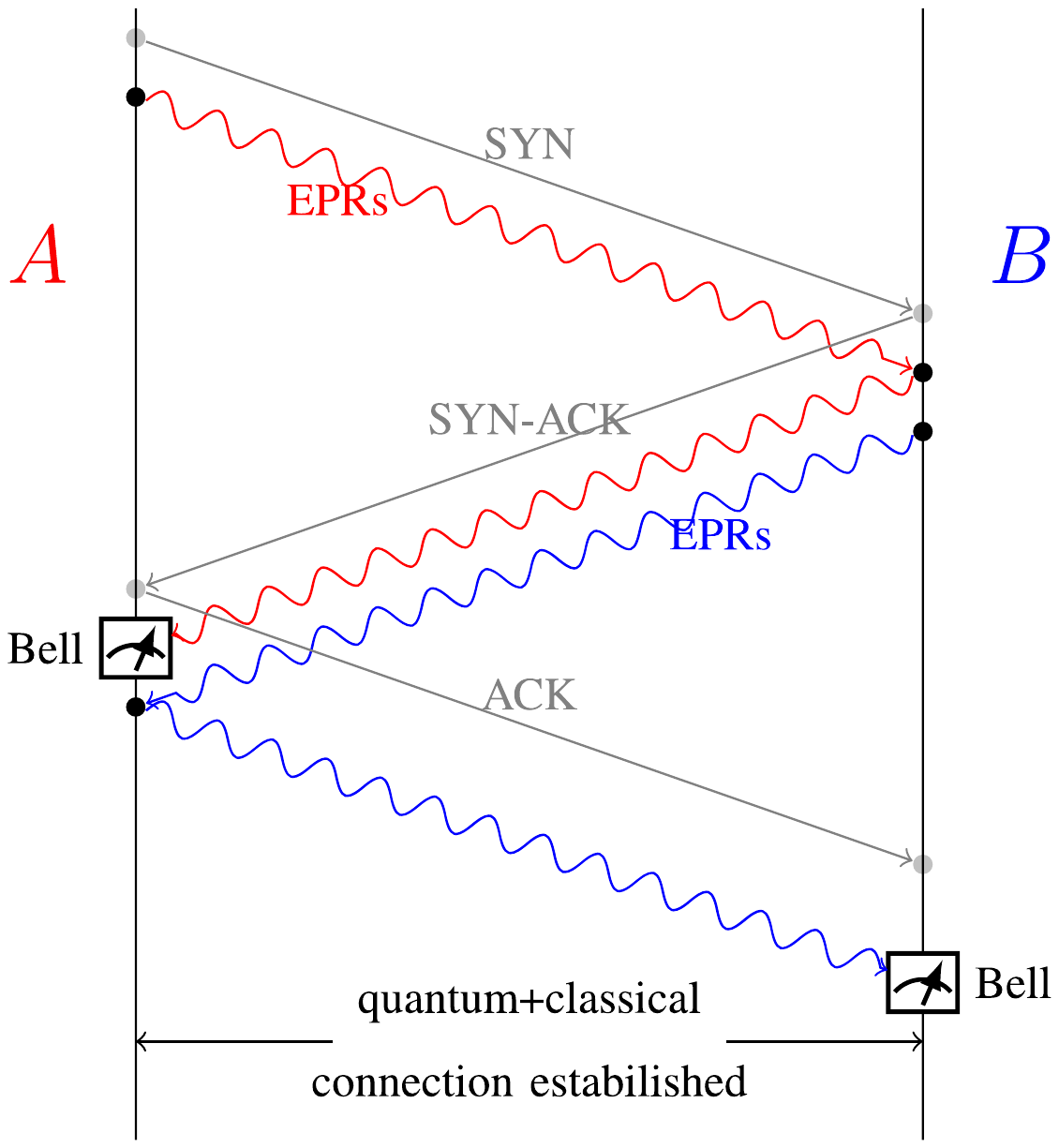} \]
	\caption{Quantum three-way handshake.
		The gray lines are classical channels. Two entangled nodes are connected by a wavy line.  } \label{fig:3}
\end{figure}
Host B first establishes a passive open, and then Host A initiates an active open. To establish a quantum connection, Host A and Host B operate as follows:
\begin{enumerate}
	\item SYN: Host A establishes $m$ local EPR pairs $\ket{\Phi^{+}}_{A_1A_2}$. Host A sends  SYN to Host B,
	together with the quantum state of $A_2$ (stored as $B_2$ by Host B)   by a {qTCP packet}.
	\item SYN-ACK: Host B receives the  {qTCP packet}. Firstly, {he applies the Pauli correction} to $B_2$, and then sends SYN+1  to Host A,  together with the quantum state of $B_2$ (stored as $A_2$ by $A$).  Host~B establishes $m$  local EPR pairs $\ket{\Phi^{+}}_{B_3B_4}$. Then Host B sends ACK to Host A, together with  the quantum state of $B_3$ (stored as $A_3$ by $A$) by a qTCP packet.
	After verifying SYN+1, Host~A performs an  $m$-fold   Bell measurement on $A_1A_2$ and checks whether the measurement outcome is $0^{2m}$.
	\item ACK: Host A sends   ACK+1, and transfers the quantum state of $A_3$ to $B_3$ of Host B.
	After verifying ACK+1, Host B performs an    $m$-fold   Bell measurement on $B_3B_4$ and check whether the measurement outcome is $0^{2m}$.
\end{enumerate}
At this point   both hosts have received an acknowledgment of the connection.
One can observe that the quantum channel between the two hosts is noiseless if and only if the distribution of EPRs is noiseless.
This protocol can accurately detect channel noises. 
For illustration, we consider the extreme case  that the entanglement is destroyed either during the teleportation from  $A_2$ to $B_2$
or from $B_2$ to $A_2$. Assume that the state of $A_1A_2$ at the end of step 2) is $\rho_{A_1A_2}$, which is not entangled (or so-called \emph{separable}). In other words, there exist probability distribution $p_i$ and quantum states $\ket{\psi_i}_{A_{1,i}}$ and $\ket{\varphi_i}_{A_{2,i}}$ such that $\rho_{A_1A_2}=\sum_i p_i \op{\psi}{\psi}_{A_{1,i}}\otimes \op{\varphi}{\varphi}_{A_{2,i}}$.
When we perform an $m$-fold Bell measurement on it, the probability of obtaining $0^{2m}$ is
\begin{align*}
&\tr\left( \langle{\Phi^{+}}\right|^{\otimes m}\rho\left|{{\Phi^{+}}}  \rangle^{\otimes m}  \right)\\
\leq& \max_i \tr\left(\langle{\Phi^{+}}|^{\otimes m}(\op{\psi_i}{\psi_i}\otimes \op{\varphi_i}{\varphi_i})|{\Phi^{+}}\rangle^{\otimes m}\right)\\
=& \max_i |\bra{{\Phi^{+}}}^{\otimes m}\ket{\psi_i\otimes\varphi_i}|^2\\
\leq& \frac{1}{2^{m}}.
\end{align*}
This probability gets smaller as the growth of $m$. 
If a noiseless quantum connection is established, our protocol reports ``Successfully connected.'' 
If the connection is  noisy,   our protocol returns ``Successfully connected.'' with a   small probability.

The classical and quantum connections for one direction are created by steps 1) and 2) and they are acknowledged. The classical and quantum connections for the other direction are created by steps 2) and 3) and they are acknowledged. Consequently a full-duplex quantum communication is established.
The detailed operations of hosts $A$ and $B$ in the quantum three-way handshake protocol are summarized in 
Protocols \ref{qR1} and \ref{qR2}, respectively.

\begin{algorithm}
	Create $2m$ qubit EPRs  $\ket{\Phi^{+}}_{A_1A_2}$\;
	Generate a random SYN\;
	Send $A_2$ together with SYN to Host B\;
	\tcc{Host B will send SYN+1, ACK and $A_2$, $B_2$ to Host A if he receives $A_1$ and SYN successfully.}
	Receive $A_2$, $B_2$ and SYN+1\;
	Measure $A_1A_2$ in the $m$-fold Bell basis\;
	\tcc{The received message is called valid if SYN+1 is consistent with the sent SYN and the measurement outcome is $0^{2m}$}
	\If{Valid message}
	{
		Send $B_2$ and ACK+1 to Host B\;
	}
	\Else
	{
		Send Abort\;
	}
	\caption{\textsf{Quantum three-way handshake (Host A)}}
	\label{qR1}
\end{algorithm}

\begin{algorithm}
	Create $2m$ qubit EPRs  $\ket{\Phi^{+}}_{B_1B_2}$\;
	Generate a random ACK\;
	Receive $A_2$ and SYN\;
	Send $A_2$ together with SYN+1, $B_2$ and  ACK+1 to Host A\;
	\tcc{Host A will send ACK+1 $B_2$ to Host B if she verifies $A_2$ and SYN+1 successfully.}
	Receive $B_2$ and ACK+1\;
	Measure $B_1B_2$ in  the $m$-fold Bell basis;\;
	\tcc{The received message is called valid if ACK+1 is consistent with the sent ACK and the measurement outcome is $0^{2m}$.}
	\If{Valid message}
	{
		ready for transmission\;
	}
	\Else
	{
		Send Abort\;
	}
	\caption{\textsf{Quantum three-way handshake (Host B)}}
	\label{qR2}
\end{algorithm}

\subsection{Data transfer}

A reason that the classical TCP works well is because  classical information can be correctly read and copied.
To handle quantum retransmission, we start with a simple question:
How to reliably send a one-qubit state $\ket{\psi}$ from Host A to Host B through a noisy quantum channel?
\noindent Usually this question is handled by using a quantum error-correcting code, which is designed for a  certain error model. Unfortunately, the channel dynamics of a quantum network varies dramatically.
For example, there are many routes connecting two endnodes and each node on a route may handle many packets simultaneously. Since the EPRs are not perfect all the time, packet loss is inevitable.    
However, we cannot afford the loss of a quantum packet without any backup.  Herein we consider a retransmission protocol based on a variant of quantum error-correcting codes--quantum secret sharing--so that at the end of the protocol we can be sure that the state is perfectly transmitted.

Our quantum retransmission protocol is given in Protocol~\ref{prot:retransmission}.
\begin{algorithm}
	\Input{$A$: register with qubits to be sent}
	Enocde $A$ by the $(2,3)$ threshold scheme and obtain $A_1A_2A_3$\;
	$k\leftarrow \mathrm{true}$\;
	\While{$k$}{
		Send $A_2$ to Host B\;
		\tcc{Host B will send valid acknowledgement to Host A if he received $B$ successfully.}
		\If{Valid acknowledged}
		{
			Send $A_3$ to Host B\;
			\If{Valid acknowledged}
			{\
				Release $A_1$\;
				$k\leftarrow \mathrm{false}$\;
				\tcc{Host B is able to recover $A$ by decoding $A_2A_3$.}
			}
			\Else
			{
				Call Quantum retransmission($A_1$)\;
			}
		}
		\tcc{If data packet is not arrived within expected time or data is damaged, Host B sends invalid acknowledgment.}
		\Else{
			Regenerate the original $A$ from current $A_1A_3$\;
			Call Quantum retransmission($A$)\;
		}
	}
	\caption{\textsf{Quantum retransmission($A$)}}
	\label{prot:retransmission}
\end{algorithm}
The kernel of the protocol is a recursive use of the $(2,3)$ threshold scheme~\cite{CGL99}.  
Note that as a initial study of this research direction, we choose the  $(2,3)$ threshold scheme for illustration of quantum retransmission.
In fact, the $(2,3)$ threshold scheme is qutrit-based (a three level quantum system); we can simply represent a qutrit by two qubits without using the fourth level.

First, a quantum message $A$ is encoded into three shares $A_1A_2A_3$ and at least two shares are required to recover the message. Then the shares $A_2$ and $A_3$ are sent sequentially, depending on the acknowledgment of the previous share. If the first share $A_2$ is acknowledged, the share $A_3$ is then sent. If $A_3$ is also acknowledged, we are done since $A_2A_3$ is enough for recovering $A$. However, if $A_2$ gets lost,
$A_1$ has to be sent carefully. 
For that purpose, $A_1$ is then encoded into three shares by the (2,3) threshold scheme
and the procedure proceeds recursively. 
Figure~\ref{fig:retransmission} illustrates these ideas.
\begin{figure*}[h]\[
	\includegraphics[width=7cm]{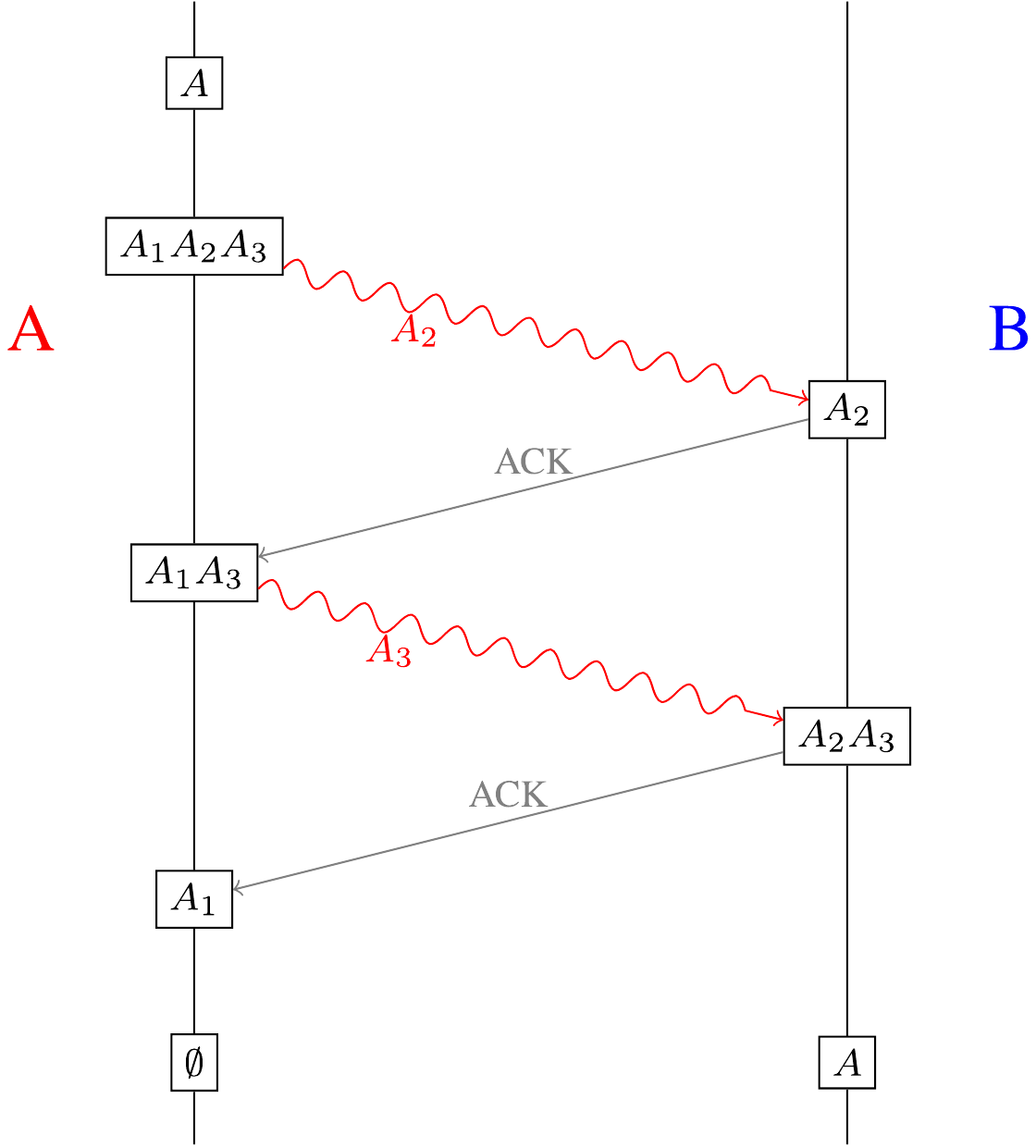}  
	\includegraphics[width=7cm]{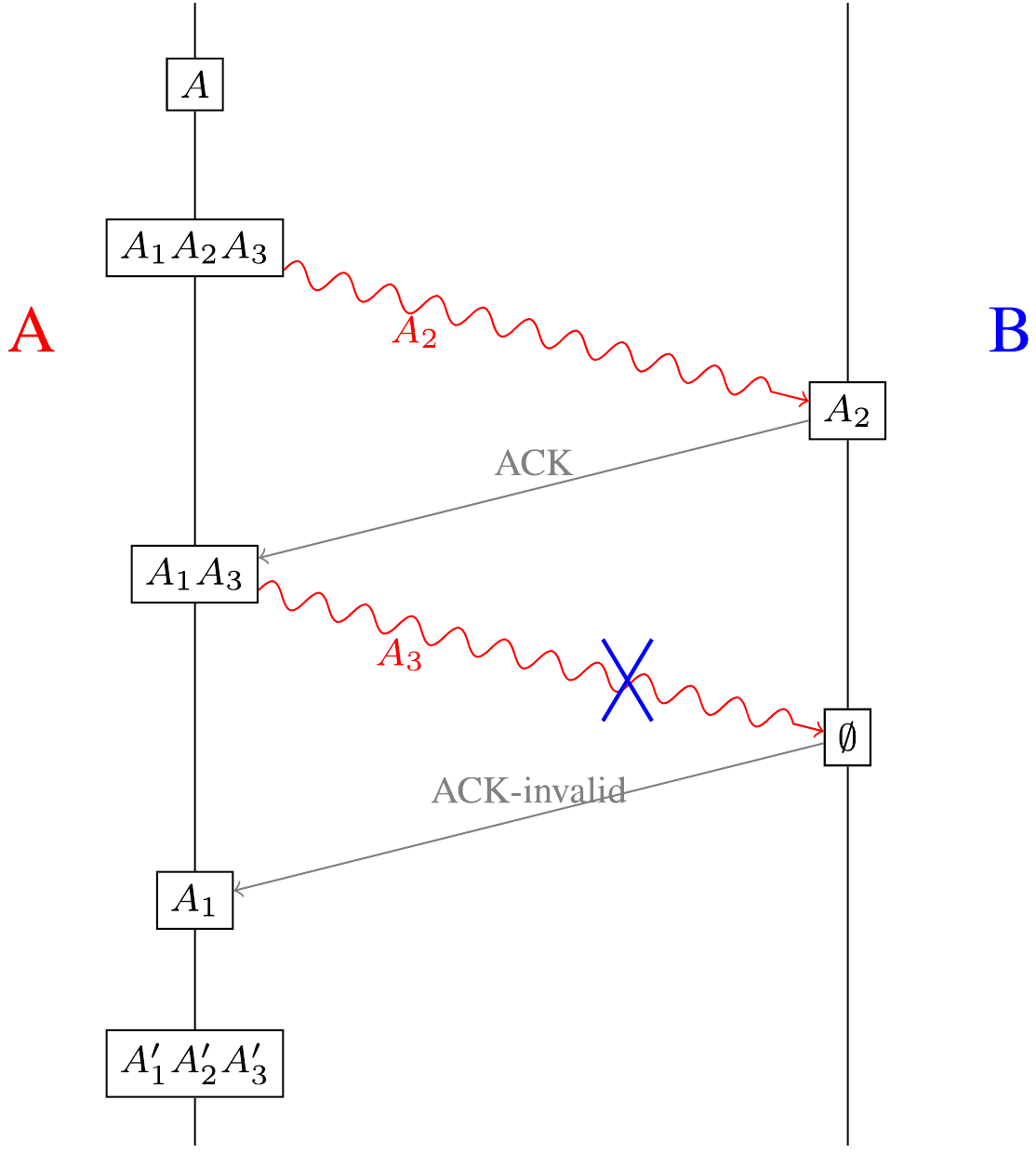}  
	\]
	\caption{ Left: Successful quantum transmission. Right: Unsuccessful Quantum transmission in the second send. Host A has to do one more recursion.
	} \label{fig:retransmission}
\end{figure*}

After a close scrutiny, one can find that the transmission succeeds if two consecutive shares are correctly received. Otherwise, either the remaining two shares have to be used to recover the first share
or the remaining one share has to be encoded into three shares.

The above argument shows that a packet loss error can be corrected. In reality, other errors can occur. For instance, each packet can be changed during transmission. We always assume that such an error can be detected through measurements on each packet, and a packet with detected errors will be discarded. Now 
we are facing a postselection issue, meaning that a packet may be measured and collapsed into some unwanted state. In the following, we show that such error can also be corrected.

Observation 1: Quantum error correction works in a tensor product manner.

Suppose that the noisy channel $\mathcal{N}$ can be corrected through the encoding channel $\mathcal{E}$ and decoding channel $\mathcal{D}$.
Then, the encoding channel $\mathcal{I}\otimes\mathcal{E}$ and decoding channel $\mathcal{I}\otimes\mathcal{D}$ can correct the noisy channel 
$\mathcal{I}\otimes\mathcal{N}$.

Observation 2: We can always assume that the input state is a pure state by adding ancilla qubits. Then, we observe that the decoding operator can still correct errors even after postselection as follows: for any $\sigma$ such that $Supp(\sigma)\subset Supp(\mathcal{N}\circ\mathcal{E}(\op{\psi}{\psi}))$, we always have
\begin{align*}
\mathcal{D}(\sigma)\propto \op{\psi}{\psi}.
\end{align*}
Combining these two observations, we know that the postselection of a new packet does not effect the quantum secret sharing scheme; the joint state between the sender and receiver still has its support lie in the code space such that the recovering operation for the packet loss still works.

{
Remark: our retransmission protocol is actually a reconstruction protocol so that the quantum package can be reconstructed at the receiver by iteratively using the recovery operation of a secret-sharing scheme.
}

Let us estimate the number of shares need to be sent. For simplicity we assume that the probability that a share gets lost in transmission is $p$
during a period of time. This can be modeled as a simple absorbing Markov chain with three states $\{0,1,2\}$, where state $i$ denotes the status of $i$-consecutive successful transmission, and a transition matrix
$$\begin{bmatrix}p &1-p & 0\\p& 0&1-p\\ 0& 0 & 1\end{bmatrix}.$$
One can show that the expected number of total shares required to send one quantum message is $\displaystyle \frac{2-p}{(1-p)^2}$. 
It can be checked that when $p<0.5$, no more than ten shares are required to send one quantum message.
When $p$ gets larger than $0.8$, the required number grows quickly.

In a less noisy channel, one can guarantee  that this scheme  succeeds within a finite number of steps.
However, a potential problem here is that the storage of a receiver for this quantum message is not unbounded.
Although Host B may need an unbounded classical data structure to maintain the status of  transmission, he has only four types of possible status:
\begin{itemize}
	\item B is {waiting for some} $A_2^{(i)}$ since {the previous} $A_2^{(i-1)}$ {is not received or valid};
	\item $A_2^{(i)}$ is received, and B is waiting for  $A_3^{(i)}$;
	\item $A_2^{(i)}$ {is received, but the corresponding} $A_3^{(i)}$ {is not valid};
	\item $A_2$ and $A_3$ {are both validly received}.
\end{itemize}
The Pseudo acknowledgment number and Pseudo Window of the qTCP packet are used for the acknowledgment of the status of Host B.

We notice that this protocol succeeds if there are two consecutive successful transmissions. Therefore, to prevent the packet size from becoming unbounded due to  potentially unbounded recursive depth, during data transmission, the sender and the receiver are required to synchronize the status of each block of original data, $A$. For example,  assume that $A_2$ is transmitted successfully but $A_3$ is not. Now the sender and receiver will notify each other that the quantum message to be sent is $A_1$. In other words, they will rename the buffer of $A$ by the buffer of $A_1$ and the packet.

\section{CONCLUSION }

We have discussed the interconnection of packet quantum network intercommunication for quantum repeater network. In particular, we proposed quantum Transmission
Control Protocols, which provide reliable quantum packet communication.
The (2,3) threshold scheme was used for illustration. One may develop  a general quantum retransmission scheme based on any quantum secret sharing scheme and find an optimal scheme with high data rate. 
One of the shortcomings of packet-switched quantum networks is that it requires low error rate of transmission, that is, it is more challenging in the hardware. This would prevent the implementation of packet-switched quantum networks in the near future.
One may consider a  quantum network  based on circuit switching for near-term devices.

The main cost of our retransmission protocol is the number of shares required to send a quantum message or the number of secret-sharing recovery operations required. Consequently, we have analyzed the expected number of shares (say $M(p)$) for a given packet loss rate $p$. Then the quantum hardware has to support a consecutive implementation of at least $2M(p)$ complete  QEC  procedures. Thus, the coherent time of a quantum memory should be much larger than the implementation time of $2M(p)$ QEC procedures.  If a (2,3) threshold scheme based on an $N$-qubit quantum code is used, roughly $2NM(p)$ EPR pairs are required. So we can estimate the size of quantum memory required for each node to be $2NM(p)$ per quantum message. These numbers may provide the minimum resource for demonstrating an experiment on quantum internet, although the scale is still beyond today’s technology.

%

The next step is to study techniques for congestion avoidance and control of quantum network protocols. In classical internet, packet switching introduces new complexities, since the packets must be re-ordered and reassembled at the destination. Also, since there are no dedicated circuits, the network links can become congested, potentially resulting in  lost packets. Quantum effects bring new challenges in developing congestion control algorithms for  quantum Transmission
Control Protocols.

Another interesting project is to produce a detailed
specification of these protocols and implement a prototype so that some initial
simulation, and in the future some experiments, with it can be performed.

	\section*{ACKNOWLEDGEMENT}
	NY thanks You Wang's generous help in the preparation of this paper and Peter Rohde for helpful communication.
	CYL thanks Kai-Min Chung for comments that improve the presentation of this paper.

	
\end{document}